\documentclass[%
 reprint,
 amsmath,amssymb,
 aps,
]{revtex4-2}

\usepackage{graphicx}% Include figure files
\usepackage{dcolumn}% Align table columns on decimal point
\usepackage{bm}% bold math
\usepackage{balance}
\usepackage{amssymb}
\usepackage{mathrsfs}
\usepackage{amsfonts} 
\usepackage{mathtools}

\usepackage{epsfig,amsmath,amsthm,bm}
\usepackage{algorithm}
\usepackage{algorithmic}

\usepackage{scalerel}

\usepackage{makecell}
\usepackage{verbatim}

\usepackage{xcolor}

\usepackage[inline]{enumitem}

\usepackage[flushleft]{threeparttable} 

\usepackage{booktabs}
\usepackage{stfloats}

\usepackage{dsfont}
\usepackage{esint}

\usepackage{url}

\usepackage{relsize}
\usepackage{subdepth}\allowdisplaybreaks

\begin{document}

\preprint{APS/123-QED}

\title{Reconfigurable-Metasurface-Aided Multi-State Generalized Polarization-Space Modulation for Next-Generation Wireless Communications}% Force line breaks with \\

%\author{John A. Hodge, Kumar Vijay Mishra, Quang M. Nguyen and Amir. I. Zaghloul
%\thanks{J. A. H. is with Bradley Department of Electrical and Computer Engineering, Virginia Tech, Falls Church, VA 22043 USA. Email: jah70@vt.edu.}
%\thanks{K. V. M. is with United States CCDC Army Research Laboratory, Adelphi, MD 20783 USA. E-mail: kumarvijay-mishra@uiowa.edu.}
%\thanks{A. I. Z. is with Bradley Department of Electrical and Computer Engineering, Virginia Tech, Falls Church, VA 22043 USA and United States CCDC Army Research Laboratory, Adelphi, MD 20783 USA. E-mail: amirz@vt.edu.}
%\thanks{J. A. H. acknowledges support from Northrop Grumman Mission Systems (NGMS), Baltimore, MD, for his thesis research. K. V. M. acknowledges support from the National Academies of Sciences, Engineering, and Medicine via Army Research Laboratory Harry Diamond Distinguished Postdoctoral Fellowship.}
%}

\author{{John} A. Hodge$^{\dag1}$,
	Kumar Vijay Mishra$^{\ddag2}$,
	Quang M. Nguyen$^{\ddag3}$,
and Amir. I. Zaghloul$^{\dag\ddag4}$}
    \affiliation{$^{\dag}$Bradley Department of Electrical and Computer Engineering, Virginia Tech, Blacksburg, VA 24061 USA\\
	$^{\ddag}$United States CCDC Army Research Laboratory, Adelphi, MD 20783 USA}
	\email{\{$^1$jah70, $^4$amirz\}@vt.edu, $^2$kumarvijay-mishra@uiowa.edu, $^4$quang.m.nguyen7.ctr@mail.mil}
	
%\author{John A. Hodge}
%\email{jah70@vt.edu.}    
%\thanks{J. A. H. acknowledges support from Northrop Grumman Mission Systems (NGMS), Baltimore, MD, for his thesis research.}
%\affiliation{Bradley Department of Electrical and Computer Engineering, Virginia Tech, Blacksburg, VA 24061 USA.}

%\author{Kumar Vijay Mishra}%
%\email{kumarvijay-mishra@uiowa.edu.}
%\thanks{K. V. M. acknowledges support from the National Academies of Sciences, Engineering, and Medicine via Army Research Laboratory Harry Diamond Distinguished Postdoctoral Fellowship.}
%\affiliation{United States CCDC Army Research Laboratory, Adelphi, MD 20783 USA.%\\This line break forced with \textbackslash\textbackslash
%}%

%\author{Quang M. Nguyen}
%\email{quang.m.nguyen7.ctr@mail.mil.}
%\affiliation{%
%United States CCDC Army Research Laboratory, Adelphi, MD 20783 USA.%\\This line break forced% with \\
%}%

%\author{Amir I. Zaghloul}
%\email{amirz@vt.edu.}
%\affiliation{Bradley Department of Electrical and Computer Engineering, Virginia Tech, Blacksburg, VA 24061 USA.}
%\affiliation{%
%United States CCDC Army Research Laboratory, Adelphi, MD 20783 USA.}


\date{\today}% It is always \today, today,

% 230 words 1,645 characters
% Needs to reduce size of abstract to 600 characters or less
\begin{abstract}
We propose a low-cost reconfigurable metasurface transmitter to implement generalized polarization-space modulation (GPSM) that has been proposed for next-generation wireless communications. Unlike conventional multiple-input multiple-output GPSM, our electromagnetically-compliant, reduced form-factor design using Barium-Strontium-Titanate film yields more than two polarization states. Full-wave simulations show our optimized design achieves 0.08-0.95 polarization conversion ratio with 5\% bandwidth at 28 GHz and bit error rates comparable or better than conventional systems.%At $15$ dB receive signal-to-noise ratio, the RMTS-based GPSM achieves $2.7$ times lower BER than a traditional MIMO quadrature amplitude modulation (QAM) system for the same SE and four times greater EE.

\end{abstract}

%Use showkeys class option if keyword
\maketitle

%\tableofcontents

\section{Introduction} \label{sec:intro}
In conventional wireless communications systems, the user messages are encoded by distinct complex vectors or codewords such as a quadrature amplitude modulation (QAM) constellation. Lately, demands\cite{qi2019outage} for high data rates in communications networks proposed for future fifth-generation (5G) and beyond are driving research in exploiting the indices of other transmission resources such as frequency subcarriers, antenna elements or subarrays (SAs), and temporal slots to allow encoding of additional information bits. This technique of \textit{index modulation} \cite{ishikawa201850,basar2016index} (IM) has recently garnered significant attention because it offers improvements in data rates, spectral efficiency (SE), and energy efficiency (EE). For example, the indices of the transmit (Tx) antennas of a multiple-input multiple-output (MIMO) system \cite{di2014spatial} and the frequency subcarriers \cite{basar2016index} are used for encoding in the basic IM forms known as spatial modulation (SM) and orthogonal frequency division multiplexing (OFDM), respectively. 

Prior works on IM focus largely on using the spatial, temporal, and frequency indices. Recent works have introduced polarization IM (PolIM) \cite{zhang2017polarization}, which employs a generalized polarization shift keying (PolarSK) modulation scheme for dual-polarization (DP) communications and leverages Tx polarization states as additional indices to achieve high SE. A further improvement in bit-error-rate (BER)  was obtained by extending \cite{henarejos20183} the classical two-dimensional (2-D) polarization modulation (PM) constellation to 3-D by mapping symbols from a sphere to respective horizontal (H-) and vertical polarizations (V-pol). Here, the indices of both polarization states and phase of a radiated electromagnetic (EM) wave were used. Similarly, a DP spatial modulation (DP-SM) \cite{zafari2016dual} considers the dimension of polarization to transmit additional information bits for correlated Rayleigh and Rician fading channels under channel estimation errors; an additional information bit is mapped to a selected (vertical or horizontal) polarization \cite{basar2017index}. 

Very recently, generalized polarization-space modulation (GPSM) \cite{henarejos20183} has been proposed as an extension of PolarSK for multiple-input multiple-output (MIMO) systems with a limited number of radio frequency (RF) chains \cite{zhang2019generalized}. It allows concurrent activation of multiple DP Tx antennas with one polarization state per antenna per time slot. The GPSM extends SM \cite{di2014spatial} through inclusion of polarization as an additional degree-of-freedom (DoF) and activating more than one antenna to increase SE. For example, if the transmit data bit is $0$ ($1$), the V-pol (H-pol) state is selected. In the conventional PolarSK system, SE is increased by employing more than two polarization states, as in \textit{slant polarization}, using a linear combination of the V- and H-pol states \cite{zhang2017polarization}. The PolarSK systems have only one DP transmit antenna so that the available DoF in the spatial domain remains unexploited. Note that the GPSM may be compared to other similar extensions of basic IMs such as generalized spatial modulation (GSM) \cite{narasimhan2015generalized} or generalized SSK (GSSK) \cite{jeganathan2008generalized}.

The 5/6G MIMO communications are envisaged to use large arrays and, consequently, require multiple RF chains. To facilitate low-cost and reduced form-factor of such systems, reconfigurable metasurface (RMTS) \cite{hodge2019reconfigurable} is currently being investigated for feasibility \cite{basar2019wireless, hodge2019reconfigurable, mishra2019reconfigurable, basar2020reconfigurable}. The RMTS is capable of controlling and manipulating EM waves through modified surface boundary conditions \cite{raeker2019compound, pfeiffer2013metamaterial, falcone2004babinet}. These surfaces are electrically thin and comprise an array of spatially-varying sub-wavelength scattering elements (or meta-atoms). Prior works demonstrate anisotropic MTS to manipulate EM wave polarization \cite{wu2019tunable, hao2007manipulating, pfeiffer2014high, zhang2009negative}. The RMTS performs signal processing and filtering operations through spatial and time modulation \cite{kwon2018nonlocal, landy2008perfect}. In particular, the introduction of RMTS as a wireless technology has given rise to the concept of smart radio environments \cite{di2019smart} that intelligently adapt to changing environments by overcoming signal attenuation from multi-path fading, blockages, and other uncontrollable interference. 
\clearpage
\onecolumngrid
%-------------------------------------------
\begin{figure*}%[ht!]
  \centering
  \includegraphics[width= 1.0 \textwidth]{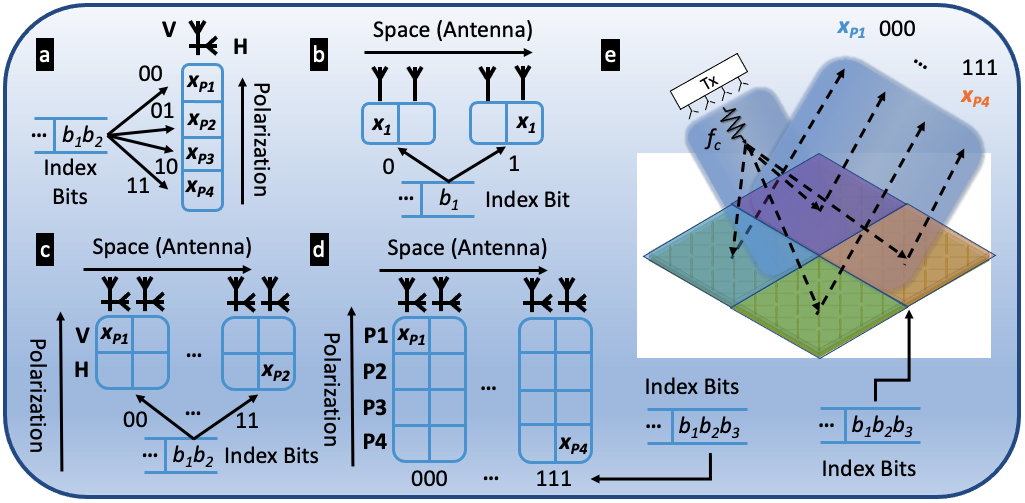}
  \caption{Schematic of the  transmit index mapping approaches for conventional (a) PolarSK \cite{zhang2017polarization} (single DP antenna with only polarization states used for mapping) (b) SM \cite{di2014spatial} (multiple antennas without any polarization diversity) (c) DP-SM \cite{zafari2016dual} (only two polarizations per antenna) (d) GPSM \cite{zhang2019generalized} (multiple polarizations per antenna), and our proposed (e) RMTS-enhanced GPSM transmitter. The user message is given by the sequence of bits ${b}_0,{b}_1,\cdots$. RMTS is illuminated with an incident carrier wave with a single polarization. The MTS inclusions are electrically tuned to reflect the modulated signal with the desired polarization state. } % \textcolor{red}{single polarization} \textcolor{blue}{should this not be multiple polarizations for our case?}
  \label{fig:systemGraphic}
\end{figure*}
%-------------------------------------------
\twocolumngrid

The RMTS applies customized transformations to radio waves based on the propagation channel and provides low-cost multi-function integrated antennas and transceivers \cite{basar2020reconfigurable}. The RMTS directly performs signal modulation on the antenna surface thereby avoiding a large number of costly and power-hungry independent RF chains and beamforming networks \cite{basar2019wireless}. However, specific implementations and EM analyses of RMTS designs of IM systems remain relatively unexamined.

In this context, our prior works \cite{mishra2019reconfigurable, hodge2019reconfigurable} consider implementing RMTS for wireless communications through careful engineering of each meta-atom \cite{hodge2019multi, hodge2019rf, hodge2019joint, hodge2021deep} to transform an incident EM wave into an arbitrarily tailored transmitted or reflected wavefront. Similar meta-atom elements are utilized to modify the radiation patterns of antennas through near-field parasitic coupling \cite{hodge2014enhancement, hodge2015utilizing}. The RMTS is designed to operate as beam-steering reflect-array  \cite{hodge2020reflective, nguyen2020circular} or transmit-array antennas  \cite{nguyen2020circular, nguyen2020design}, which may be polarization agnostic \cite{hodge2020reflective}. In \cite{hodge2020intelligent}, we showed RMTS implementations of frequency-domain IM (FD-IM) techniques, such as OFDM-IM. Recently, we introduced RMTS-aided media-based modulation \cite{hodge2020media} to improve the bit error rate (BER) using channel as a DoF.

In this paper, we introduce the RMTS implementation of GPSM %in place of a DP phased array antenna for GPSM, we %propose its RMTS implementation. %Through reconfigurable polarization control of the RMTS and adaptive compensation, we optimize the polarization of Tx signals based on the characteristics of the propagation channel to maximize the signal-to-noise ratio (SNR) at the receiver (Rx). This paper 
to achieve a reduced form-factor that is optimized to provide maximum polarization tunability. Further, unlike the conventional DP phased array system of GPSM \cite{zhang2019generalized} that requires a MIMO Tx antenna with multiple RF chains and produced only two polarization states, our proposed design requires only single RF feed source and yields more than two polarizations. This reduces the cost and RF complexity of the transceiver. We verify the polarization conversion performance using finite-element method (FEM) full-wave EM simulations, which are later used to provide realistic co-polarization and cross-polarization levels while evaluating BER at the receiver (Rx). Figure~\ref{fig:systemGraphic} compares our RMTS-assisted GPSM scheme with PolarSK, SM, DP-SM, and GPSM. Table~\ref{tbl:comparison} quantitatively summarizes these systems.

\section{RMTS Polarizer}
%\textcolor{red}{Change the title. It is too general.}
Previous works on MTS-based polarization conversion \cite{grady2013terahertz, cheng2014ultrabroadband, fang2017design, ma2014broadband} and reconfiguration \cite{yang2016programmable} use varactor or PIN diodes as the tunable element \cite{tian2019reconfigurable, tao2017reconfigurable, chen2019dual, ratni2017electronic, yang2016programmable}. In this paper, in place of a diode, we employ MTS with a tunable Barium Strontium Titanate (BST) substrate to achieve polarization modulation. The BST is a ferroelectric material with tunable capacitance that is suitable for RF components due to its producibility, low-cost, and high speed characteristics \cite{tombak2002tunable}. %\textcolor{red}{This tunable material is easy to integrate with the meta-atom design and, unlike a diode, does not require discrete components or a matching network at each element.}

The polarization conversion principle of our RMTS polarizer design is as follows. The electric and magnetic resonant features of the unit cell allow tuning of the reflected signal between two orthogonal polarization states. 
\clearpage
\onecolumngrid
%----------------------------------------------------
\begin{table*}[ht!]
\centering
\caption{Comparison of the proposed method with the state-of-the-art}
\label{tbl:comparison}
\begin{threeparttable}
\begin{tabular}{c|c|c|c|c|c|c|c}
\hline
 \textbf{\thead{Modulation} } & \textbf{\thead{Total Tx \\ Antennas}} & \textbf{\thead{Activated Tx  \\ Antennas}} & \textbf{\thead{Polarization \\ States \\ ($P$)}} & \textbf{\thead{Polarization \\ Angles \\ ($K$)}} & \textbf{\thead{Data Rate\tnote{a}\; ($R$) \\ (bits/s/Hz)}} &   \textbf{\thead{RF \\ Chains \\ ($N_{\textrm{RF}}$)}} &   \textbf{\thead{Dynamic \\ Radiation \\ Patterns}} \\ \hline\hline
  PolarSK \cite{zhang2017polarization} & 1 & 1 &  $\ge2$ & $\ge1$ & $\log_{2}(P)$ \vspace{2pt}& 1 & No \\ \hline
  SM \cite{di2014spatial} &  $N_{\mathrm{T}} \ge 2$ & $N_{\mathrm{RF}} < N_{\mathrm{T}}$ & 1 & 1 & $\log_{2}(N_{T}) + \log_{2}(M) $ &   $N_{\mathrm{RF}}$ \vspace{2pt}& No \\ \hline
  DP-SM \cite{zafari2016dual} &   $N_{\mathrm{T}} \ge 2$ & $N_{\mathrm{RF}} < N_{\mathrm{T}}$ &  2 & 1 & $\log_{2}(P N_{T} M)$ \vspace{2pt}&   $N_{\mathrm{RF}}$ &  No \\ \hline
  GPSM \cite{zhang2019generalized} &   $N_{\mathrm{T}} \ge 2$ & $N_{\mathrm{RF}} < N_{\mathrm{T}} $ & $\ge 2$ & $\ge1$ & $\left \lfloor{ \log _{2}\left ({\begin{smallmatrix} N_{\mathrm {T}} \\ N_{\mathrm {RF}} \end{smallmatrix}}\right )}\right \rfloor +N_{\mathrm {RF}}\lfloor \log _{2}(P)\rfloor$ & $N_{\mathrm{RF}}$ \vspace{2pt}& No \\ \hline
  This paper & $N_{\mathrm{T}} \ge 2$ & $N_{\textrm{SA}} < N_{\mathrm{T}}$ & $\ge2$ & $\ge1$ & $\left \lfloor{ \log _{2}\left ({\begin{smallmatrix} N_{\mathrm {T}} \\ N_{\mathrm {SA}} \end{smallmatrix}}\right )}\right \rfloor +N_{\mathrm {SA}}\lfloor \log _{2}(P)\rfloor$ \vspace{2pt}& 1 & Yes \\ \hline
\end{tabular}
 \begin{tablenotes}[para]
 \item[a] For various notations used here, we refer the reader to the supplement on GPSM signal model.
 \item[b] The function $\lfloor\cdot\rfloor$ yields the greatest integer smaller than or equal to the argument. \textcolor{red}{
 %(1) In the data rate of SM, how does M enter the formula? In SM, there is no polarization. So, what does M represent there? Is it the same quantity as our paper? \\ 
 %(2) Similarly, the data rate for DP-SM has the product $PM$. In GPSM, $P=M^2K$. Is M related to P in DP-SM? Does it have the same meaning as M in GPSM? \\
 %(3) In GPSM, you write that $N_{RF} = N_T$. So, in the data rate formula, what does it mean by $N_T$ choose $N_{RF}$? There is no choice there because $N_T$ choose $N_T$ is simply 1. \\ 
 %(4) In GPSM data rate formula, shouldn't $N_{\mathrm {RF}}\lfloor \log _{2}(P)\rfloor$ become $N_{\mathrm {T}}\lfloor \log _{2}(P)\rfloor$? \\ 
 %(5) In GPSM, P=2. But $P=M^2K$. So, what are the values of M and K here? The only integer solutions are M=1, K=2, which means these values are fixed. Does this make sense? \\
 %(6) For `This paper', please check if it makes sense to have $N_{SA}=1$. This is possible because we write $N_{SA} < N_T$ and $N_T$ could be just 2 (second column entry is greater than or equal to 2) while the third column entry is strictly smaller than $N_T$. 
 }
 \end{tablenotes}
 \end{threeparttable}
\end{table*}
%----------------------------------------------------
%----------------------------------------------------
\twocolumngrid

\noindent Dipoles \cite{grady2013terahertz} and v-shaped \cite{cheng2014ultrabroadband} resonators rotated $45$ degrees relative to the electric ($E$) field of the incident wave $\mathbf{E}^{i}$ are used to realize linear polarization conversion metasurfaces. The unit vectors for the Cartesian $x$-, $y$-, and $z$- axes are $\hat{\mathbf{x}} = \frac{\mathbf{x}}{|\mathbf{x}|}$, $\hat{\mathbf{y}} = \frac{\mathbf{y}}{|\mathbf{y}|}$, and $\hat{\mathbf{z}} = \frac{\mathbf{z}}{|\mathbf{z}|}$. The incident wave is 
\begin{equation}
    \mathbf{E^{i}}(z) = \hat{\mathbf{x}} E_{0}e^{-\mathrm{j}k_{z}z},
\end{equation}
where $\hat{x}$ is the unit vector in the $x$-direction, $E_{0}$ is the $E$-field magnitude in volts per meter (V/m) and $k_{z}$ is the wavenumber for the wave propagating in the $-\hat{z}$ direction. The $E$-field of the wave reflected by the MTS is decomposed into two orthogonal components $u$ and $v$ such that $\mathbf{E}_{u} = \mathbf{E}_{x} + \mathbf{E}_{y}$ and $\mathbf{E}_{v} = \mathbf{E}_{y} - \mathbf{E}_{x}$. Denote the four reflected polarization components by $r_{xx} = {|\mathbf{E}_{x}^{r}| }/{|\mathbf{E}_{x}^{i}|}$, $r_{xy} = {|\mathbf{E}_{x}^{r}|}/{|\mathbf{E}_{y}^{i}|}$, $r_{yx} = {|\mathbf{E}_{y}^{r}|}/{|\mathbf{E}_{x}^{i}|}$, and $r_{yy} = {|\mathbf{E}_{y}^{r}|} / {|\mathbf{E}_{y}^{i}|}$. The incident and reflected waves are $\mathbf{E}^{i}$ and $\mathbf{E}^{r}$, respectively. The subscripts $x$ and $y$ indicate the $\hat{\mathbf{x}}$ and $\hat{\mathbf{y}}$ polarized components, respectively, of the field. The resultant reflected $E$-field ($\mathbf{E}^{r}$) is the vector sum of the components 
\begin{align}
    \mathbf{E}^{r} = E^{r}_{x}\hat{\mathbf{x}} + E^{r}_{y}\hat{\mathbf{y}}.
\end{align}

The magnitudes of $\mathbf{E}^{i}$ and $\mathbf{E}^{r}$ are related by% \textcolor{red}{the subscript changes to superscript in the equation below}
\begin{equation} \label{eqn:rRelation}
    \begin{bmatrix}
        |\mathbf{E}_{x}^{r}| \\
        |\mathbf{E}_{y}^{r}| 
    \end{bmatrix}
    =
    \begin{bmatrix}
        r_{xx} & r_{xy}  \\
        r_{yx} & r_{yy}
    \end{bmatrix}
    \begin{bmatrix}
        |\mathbf{E}_{x}^{i}| \\
        |\mathbf{E}_{y}^{i}| 
    \end{bmatrix}
    .
\end{equation}
Define the linear-to-linear cross-polarization conversion ratio (PCR) of a $y$-polarized wave as
\begin{equation}
    \textrm{PCR} = \frac{r_{xy}}{r_{yy} + r_{xy}}. \label{eqn:pcr}
\end{equation}
Given an incident $y$-polarized wave, the objective of RMTS polarizer is to create a tunable change of polarization states with a PCR between $0$ and $1$. This implies that the MTS polarizer controls the proportion of $\mathbf{E_{t}}$ between the orthogonal $x$ and $y$ polarizations. We realize the polarization states for GPSM by controlling the PCR.

The effective impedance $Z$ of the meta-atom is modeled using an effective RLC circuit as
\begin{equation}
    Z = R_{\textrm{eff}} + \mathrm{j}\left(\omega L_{\textrm{eff}}-\frac{1}{\omega C_{\textrm{eff}}}\right) = R_{\textrm{eff}} + \mathrm{j}X,
\end{equation}
where $R_{\textrm{eff}}$ is the effective resistance, $\omega = 2\pi f$ is the angular frequency, $f$ is the carrier frequency of the signal, $L_{\textrm{eff}}$ is the effective inductance produced by current on the metallic strips, $C_{\textrm{eff}}$ is the effective capacitance between metallic strips and the groundplane, and $X$ is the reactance. The impedances of the orthogonal polarization components $\mathbf{E}_{x}$ and $\mathbf{E}_{y}$ are $Z_{x} = R_{x} + \mathrm{j} X_{x}$ and $Z_{y} = R_{y} + \mathrm{j} X_{y}$, respectively. To realize the RMTS polarizer, we choose a meta-atom geometry that provides tunable $X_{x}$ and $X_{y}$ values to produce orthogonal polarization responses. Our design using a tunable BST-loaded substrate changes the effective dielectric constant $\epsilon_{r}$ of the meta-atom substrate. The meta-atom's effective capacitance $C_{\textrm{eff}}$ is a function of $\epsilon_{r}$. Subsequently, the reactance relates to $\epsilon_{r}$ as $X \sim 1 / (C_{\textrm{eff}}(\epsilon_{r}))$. % \textcolor{red}{how is this related to the previous sentence where you say we want to tune the reactances?}

The BST-loaded substrate offers tunable permittivity $\epsilon_{r}$ by applying an external bias voltage $v_{b}$ \cite{tombak2002tunable}. Using a BST thin film between two Duroid ($\epsilon_{r}=2.3$) slabs, we create a tunable composite dielectric substrate with an effective $\epsilon_{r}(v_{b})$ ranging from $3.0-5.0$ \cite{nguyen2020circular}. By controlling $\epsilon_{r}$, we also regulate $C_{\textrm{eff}}$ and hence $X$ for each polarization within tunable ranges. By adjusting the effective dielectric constant of the meta-atom substrate, we control the electric resonance of two orthogonal dipoles.

The response of the multilayer RMTS polarizer is theoretically understood using a multilayer dielectric slab model \cite{orfanidis2002electromagnetic}. When an incident wave $\mathbf{E}^{i}_{y}$ is reflected off the MTS, the scattering response for each polarization \cite{grady2013terahertz, tao2017reconfigurable} is obtained using a transfer matrix method (TMM). Using boundary conditions, the reflection $r$ and transmission $t$ %\textcolor{red}{are you using r and t for coefficients here? If yes, then it is not a good notation because you used r and t for indices earlier. As I mentioned in the next section, you could instead use $\Gamma$ and $\Tau$.} 
coefficients of forward and backward propagating waves at each substrate interface are summed up to compute the total reflected electric field \cite{ma2014broadband}. Each metallic structure layer is treated as a zero-thickness impedance sheet where the coefficients are obtained through numerical EM simulations. From this model, we gain an intuitive understanding of multilayer meta-atom function as a tunable polarizer. %\textcolor{red}{what is the conclusion of this para? where are the results?}

\clearpage
\onecolumngrid
%-------------------------------------------
\begin{figure*}[ht!]
  \centering
  \includegraphics[width=1.0\textwidth]{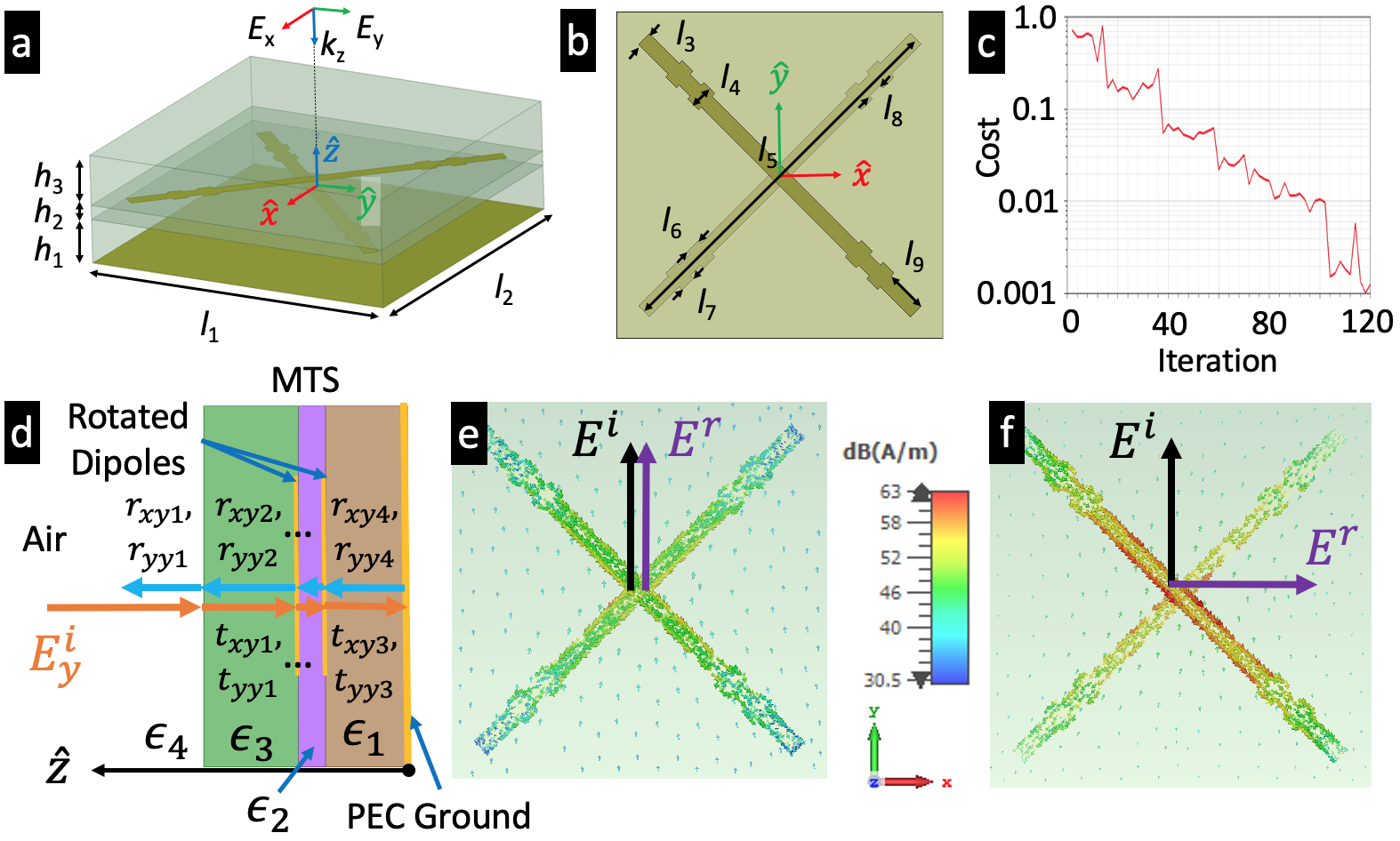}
  \caption{(a) Isometric and (b) top-down views of our polarization conversion MTS unit-cell (meta-atom) with three layers including two orthogonal metallic dipoles. The Cartesian coordinates are shown in the $XYZ$ space. The meta-atom dimensions are: $l_1 = l_2 = 2.046$ mm, $h_{1} = 0.284$ mm, $h_{2} = 0.093$ mm, $h_{3} = 0.327$ mm, $l_{3} = l_{6} = l_{7} = l_{8} = 0.096$ mm, $l_{4} = 0.146$ mm, $l_{5} = 2.413$ mm, and $l_{9} = 0.267$ mm. (c) Results of meta-atom optimization to maximize the tunable PCR range at $f = 28$ GHz. (d) Diagram of the material stack-up for the MTS in the $\hat{z}$-direction. The effective relative dielectric constants are $\epsilon_{1} = 3.5$, $\epsilon_{2} = 2.2$, $\epsilon_{3}$ is tuned between $3.0-5.0$, and $\epsilon_{4} = 1.0$. For an incident wave $\mathbf{E}^{i}_{y}$, the transmission $t$ and reflection $r$ coefficients between each material layer are shown for the $xyh$ and $yyh$ polarization components (Eqn.~\ref{eqn:rRelation}) where $h$ is the dielectric layer indices. (e-f) The surface current distribution and the polarization of the corresponding incident ($\mathbf{E}^{i}$) and reflected ($\mathbf{E}^{r}$) signals when (e) $\epsilon_{3} = 3.0$ and (f) $\epsilon_{3} = 5.0$. % \textcolor{red}{same issue with the use of r and t here. See my comment in the text}
  }
  \label{fig:unitCellDim}
\end{figure*}
%-------------------------------------------
\twocolumngrid

\section{Reflective Polarization Conversion} \label{sec:metaAtomDesign}
We design a reflection-type MTS by incorporating constituent dielectric layer materials for controlling the polarization. The proposed design (Fig.~\ref{fig:unitCellDim}) comprises three layers: the first consists of an I-shaped structure placed on top of a dielectric layer ($\epsilon_{1} = 3.5$) with metal backing; the second is a dielectric spacer ($\epsilon_{2} = 2.2$); and the third has an I-shaped structure placed on top of a constituent material and oriented $90^{\circ}$ with respect to the first layer. 

The physical dimensions of our meta-atom design are shown in Fig.~\ref{fig:unitCellDim}a-b. We maximized the tunable PCR range of our RMTS at $f = 28$ GHz using ANSYS HFSS's gradient-based quasi-Newton optimizer. The orthogonal reflection coefficients $r_{xy}(\epsilon_{3})$ and $r_{yy}(\epsilon_{3})$ are functions of the tunable substrate $\epsilon_{3}$ (Fig.~\ref{fig:unitCellDim}e-f). We optimized the dimensions of our meta-atom using design variables $h_{1}$, $h_{2}$, $h_{3}$, $l_{4}$, and $l_{5}$ (Fig.~\ref{fig:unitCellDim}a-b) over $120$ full-wave simulation iterations (Fig.~\ref{fig:unitCellDim}c) to achieve the maximum tunable PCR range subjected to the constraints%equally weighted the four constraints, %\textcolor{red}{where did you define $r_{xx}$ as a function of $\epsilon$ earlier? Also, the diagram has an additional numeral subscript to $r_{xx}$} \textcolor{red}{you must write the expression of this cost function}
\begin{align*}
    &|r_{yy}(\epsilon_{3} = 3.0)| \geq 0.95 \\
    &|r_{xy}(\epsilon_{3} = 3.0)| \leq 0.05 \\
    &|r_{yy}(\epsilon_{3} = 5.0)| \leq 0.95 \\
    &|r_{xy}(\epsilon_{3} = 5.0)| \geq 0.05.
\end{align*}

%-------------------------------------------
\begin{figure}[ht!]
  \centering
  \includegraphics[width=1.0\columnwidth]{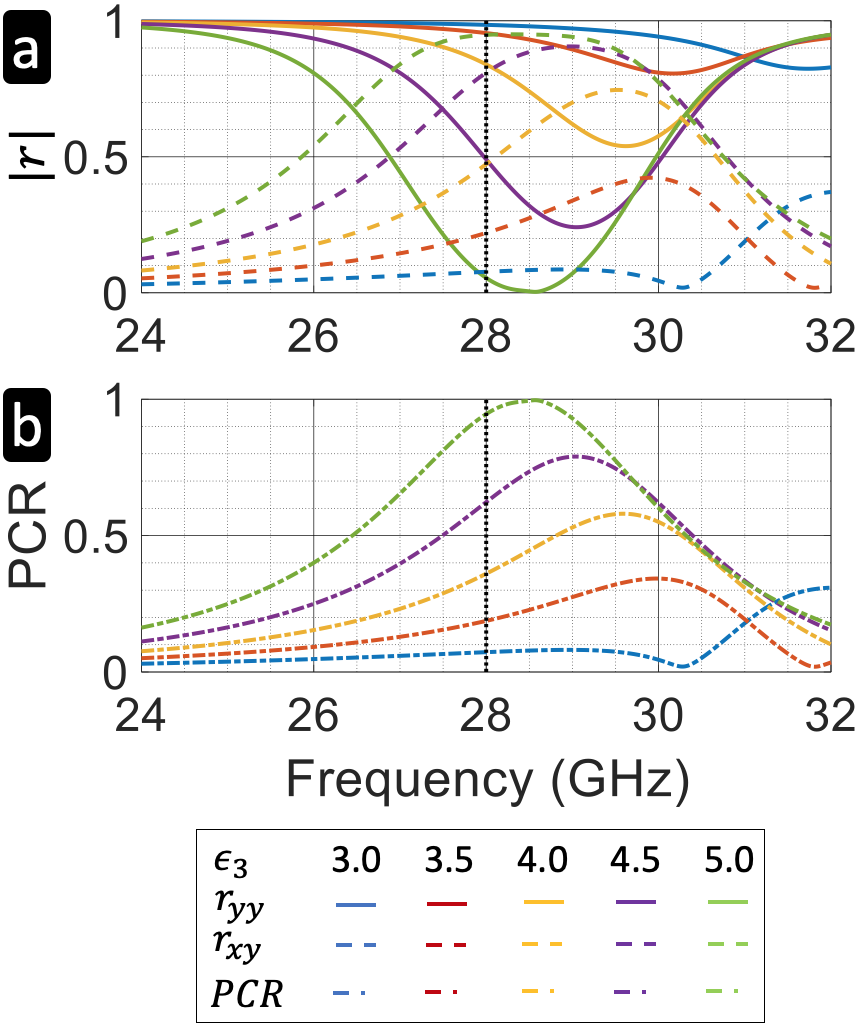}
  \caption{(a) The reflection coefficient amplitudes $|r|$ and (b) PCR over frequency from $f = 24-32$ GHz for our meta-atom in different reconfigurable tuning states as obtained in HFSS EM simulations. The effective dielectric constant for the BST-loaded spacer ($\epsilon_{3}$) is tuned between the value of $3.0$ to $5.0$ in steps of $0.5$. For $\epsilon_{3} = 3.0$, more than $95\%$ of the power from the reflected signal ($r_{yy}$) is co-polarized with the incident wave from the RF feed. For $\epsilon_{3} = 5.0$,  more than $95\%$ of the power from the reflected signal ($r_{xy}$) is converted into the orthogonal cross-polarization. The vertical dotted line denotes the design frequency of $f = 28$ GHz.
  % \textcolor{red}{Change $\Gamma$ to $r$.} 
  }
  \label{fig:reflectionEMsim}
\end{figure}
%-------------------------------------------

%The resulting cost function is 
%\begin{align*}
%    \underset{h_{1}, h_{2}, h_{3}, l_{4}, l_{5}}{\text{minimize}} &f(h_{1}, h_{2}, h_{3}, l_{4}, l_{5})  = \sum_{i=1}^{4} f_{i}(h_{1}, h_{2}, h_{3}, l_{4}, l_{5}) \\
%    &f_{1} = (0.95)^{2} - (|r_{yy}(\epsilon_{3} = 3.0)| + n_{0})^{2} \\
%    &f_{2} = (|r_{xy}(\epsilon_{3} = 3.0)| + n_{0})^{2} - (0.05)^{2} \\
%    &f_{3} = (|r_{yy}(\epsilon_{3} = 5.0)| + n_{0})^{2} - (0.05)^{2} \\
%    &f_{4} = (0.95)^{2} - (|r_{xy}(\epsilon_{3} = 5.0)| + n_{0})^{2}, \\
%\end{align*}
%where the cost function noise $n_{0} = 0.0001$ is the approximate error term. %\textcolor{red}{The Quasi-newton optimizer minimizes the cost function to achieve the desired RF performance (Fig.~\ref{fig:unitCellDim}c).}

%\textcolor{blue}{this sentence does not read correct. Where is the cost function? The equations above are inequality constraints, not a function.} 
%\textcolor{red}{Fig. 4 is referred earlier than other subfigures of Fig. 3. This is not a good way to reference. You could possibly just add Fig. 4 as a subfigure of Fig. 3.}

The orthogonal metallic dipoles are separated by height of $h_{2}$ in the $\hat{z}$-direction. Figure~\ref{fig:unitCellDim}d shows the material stack-up in the $\hat{z}$-direction. % \textcolor{red}{why not use $\Gamma$ for r earlier also? For transmission coefficient, you can use $\Tau$}
The polarization states of a linearly polarized incident wave with normal and oblique (from $-30^{\circ}$ to $+30^{\circ}$ with respect to $z$-axis) incidence are manipulated as desired after reflected by anisotropic MTS. %\textcolor{blue}{Repetition. You already mentioned these lines earlier.} \textcolor{red}{This is achieved by controlling the permittivity of the BST film embedded in the dielectric layer as a constituent material ($\epsilon_{3}$ = 3.0-5.0). The permittivity of BST film is tuned by applying voltage bias.} \textcolor{blue}{Too much going back and forth between subfigures of Fig. 3. Why is the following line here instead of the earlier para where you described all the layers.} \textcolor{red}{The layers of the MTS polarizer are shown in Fig.~\ref{fig:unitCellDim}a and Fig.~\ref{fig:unitCellDim}c.} The area above the MTS is free space (air) with $\epsilon_{4} = 1.0$.
The key feature of this design is that it allows the conversion of different reflected polarization states such as linear, circular, and cross-polarized waves. Furthermore, the effect of the oblique incident wave with respect to polarization conversion efficiency is also governed by tuning the permittivity of BST film. The present design works for both normal and oblique (from $-30$ to $+30^{\circ}$ with respect to $z$-axis) incidences. At the operating frequency $f_{c} = 28$ GHz, the meta-atom has electrical dimension of $\lambda_{0}/6 \times \lambda_{0}/6 \times \lambda_{0}/12$. %cost function criteria. 

%Figure~\ref{fig:reflectionEMsim} shows the optimized reflection coefficient $|r|$ and PCR performance over a broad tunable range to meet the design constraints.

%\textcolor{blue}{Lot of repetition. Already said these things in earlier sections. Either remove from there or here. Or keep only the new conclusions here.} \textcolor{red}{The polarization conversion nature of this surface stems from our three-layer design. The reflection phases between the first and third layers are modulated independently by tuning BST film's permittivity, which provides the reconfigurability for controlling different polarization states of reflected waves.}

%To achieve good signal isolation in a communications system, we limit the design to only linearly polarized and cross-polarized waves. \textcolor{red}{But Table I said we are doing more than 2 states} In other words, the proposed MTS converts linearly polarized incident wave into either vertical or horizontal polarization. 

We used Ansys HFSS software to perform full-wave EM simulations of our meta-atom design. We also simulated the unit cell in CST Microwave Studio as an additional verification. Figures~\ref{fig:unitCellDim}e-f show the surface current distribution for our meta-atom in the co- and cx-polarized states. The bias value of the BST-loaded substrate changes the induced current distribution, resonance, and coupling of our anisotropic meta-atom allowing for polarization rotation. Our MTS rotates polarization of reflected signal from $\mathbf{E}^{r}_{y}$ to $\mathbf{E}^{r}_{x}$ by tuning $\epsilon_{3}$.

The co- and cross-polarized (cx) reflection \textcolor{black}{coefficient magnitudes ($|r|$)} and PCR of our reconfigurable polarization meta-atom in different tuning states are shown in Fig.~\ref{fig:reflectionEMsim}. Changing the bias voltage of the tunable BST material reflects a signal with either $90\%$ linear co- or cx-polarized magnitude at $f_{c} = 28$ GHz depending on the tuning state. We perform polarization modulation using this tunable mechanism to transmit either a vertical, horizontal, or composite slant polarized signal from the MTS. This MTS design provides a reconfigurable PCR between values of $0.08$ to $0.95$ (Fig.~\ref{fig:reflectionEMsim}b). This design maintains a PCR greater than $0.9$ over a frequency bandwidth (BW) from $27.8-29.2$ GHz ($5.0\%$ BW). If the required tunable PCR range is relaxed, the operational BW is broadened significantly. When $\epsilon_{3} = 3.0$, the device maintains a PCR $\leq 0.1$ across $24$-$30.7$ GHz.

\section{Numerical Experiments} 
\label{sec:numExp}
We performed numerical experiments to evaluate the performance of the proposed RMTS-assisted GPSM MIMO system. Throughout the experiments, we use the RMTS Tx, $N_{R} = 4$, and $K_{c} = 10$. We employ the maximum likelihood (ML) detector for symbol detection. We analyze the performance by varying $N_{T}$, $N_{\textrm{SA}}$, $M$, and $K$ with signal-to-noise ratio (SNR) ranging from $0$ to $25$ dB. In particular, we investigate GPSM cases with $N_{\textrm{SA}} = 1$ or $2$. We set
\begin{align}
    \epsilon_{k} = 
    \begin{dcases}
    \frac{\pi}{4},\; &K=1,\\
    \left\{\frac{\pi}{4}, \frac{\pi}{8}\right\}, \;& K = 2.
    \end{dcases}
\end{align}

\begin{figure*}
  \centering
  \includegraphics[width=1.0\linewidth]{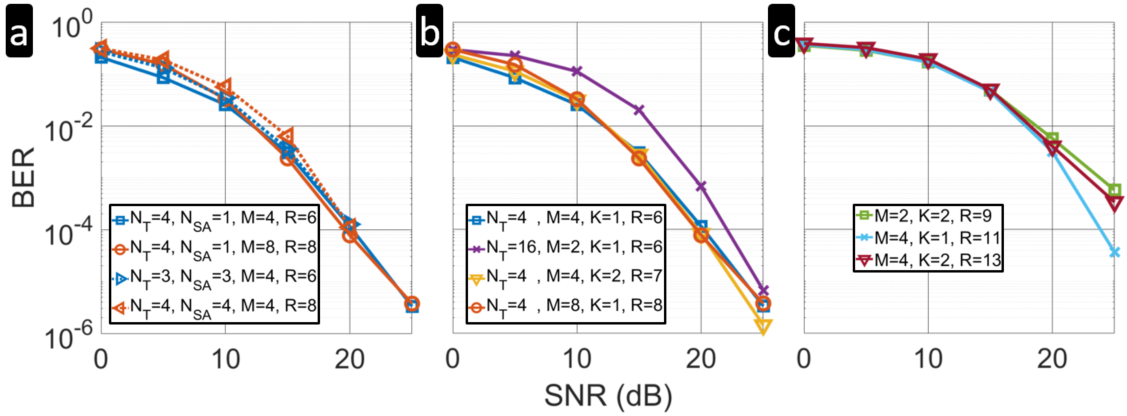}
  \caption{BER versus SNR for RMTS-assisted GPSM with $N_{R} = 4$ antennas at the Rx. (a) Performance comparison of GPSM with $K=1$ (solid lines) and traditional MIMO-QAM (dashed lines) for cases with equal SE. Performance of selected GPSM cases when (b) $N_{\textrm{SA}}=1$ and (c) $N_{\textrm{SA}}=2, N_{T}=5$. The $R$ is the SE in bits per channel use (bpcu). %\textcolor{red}{Make sure you show the effect on the performance by using subarrays and polarization selection too.}
  }
  \label{fig:plot_BER_GPSM}
\end{figure*}

Figure~\ref{fig:plot_BER_GPSM}a compares the BER of GPSM with the conventional MIMO quadrature amplitude modulation (QAM). The RMTS-assisted GPSM performs as well as or better than the traditional MIMO-QAM systems when SE or $R$ are similar. Our numerical experiments show that our RMTS-based GPSM transmitter achieves a $2.7$ times lower BER than a traditional MIMO-QAM transmitter at SNR$=15$ dB for the same SE and with four times greater EE, since only one subcarrier is used at each signaling interval rather the four. In this example, we utilize only one subcarrier at each signaling interval rather than four simultaneous subcarriers as in MIMO-QAM. The RMTS-based GPSM system uses less activated $N_{\textrm{SA}}$ and RF chains resulting in a superior EE. % \textcolor{red}{while SE is defined in eq. (9), where is EE defined?}

We compare the performance of selected GPSM cases when $N_{\textrm{SA}}=1$ (Fig.~\ref{fig:plot_BER_GPSM}b) and $N_{\textrm{SA}}=2$ (Fig.~\ref{fig:plot_BER_GPSM}c). Our BER results show that an increase in $M$ tends to be more effective than an increase in $N_{T}$ while keeping other parameters the same. When $N_{\textrm{SA}} = 1$, increasing $K$ from $1$ to $2$ enhances $R$ by a bit without sacrificing the BER performance (Fig.~\ref{fig:plot_BER_GPSM}b). The SE is also increased by using $N_{\textrm{SA}} = 2$ with $N_{T} = 5$ (Fig.~\ref{fig:plot_BER_GPSM}c). The increased throughput is at the cost of deteriorated BER. By utilizing the polarization states as a DoF, the GPSM system encodes additional information using the polarization states. %Future performance is further increased by further optimizing the cross-polarization performance of the RMTS transmitter.

\clearpage
\onecolumngrid
%-------------------------------------------
%-------------------------------------------
\twocolumngrid

\section{Summary}
\label{sec:conclusion}
We presented RMTS design to simplify wireless communications architectures to support the next-generation IM waveforms that employ polarization as a DoF, such as PolarSK and GPSM. We demonstrated that our BST-based RMTS-assisted GPSM is a promising solution compared to the traditional MIMO-QAM systems when higher SE and EE are required. Our optimized RMTS design provides PCR between values of $0.08$ to $0.95$ for tunable polarization conversion at $28$ GHz. Our numerical simulations show an appreciable BER improvement of RMTS-based GPSM over conventional MIMO-QAM transmitters at SNR values of $\geq 15$ dB and provides four times high EE for the same SE. 

%\textcolor{red}{write some more specific conclusions or quantitative results of our work. The sentences that I commented are general/not GPSM-related.} 
%The MTS-based integrated antenna-transceiver architecture provides compelling advantages of low RF complexity, cost, and power consumption compared to current phased array antennas and heterodyne transceiver architectures. %Additionally, RMTS complexity and cost advantages over traditional phased arrays scale with increasing aperture size. This makes direct modulation RMTS transmit arrays a compelling solution for IM, massive MIMO, and shared-aperture multi-beam antennas in 5G and 6G wireless networks.

\begin{acknowledgments}
J. A. H. acknowledges support from Northrop Grumman Mission Systems (NGMS), Baltimore, MD, for his thesis research. K. V. M. acknowledges support from the National Academies of Sciences, Engineering, and Medicine via Army Research Laboratory Harry Diamond Distinguished Postdoctoral Fellowship.
\end{acknowledgments}

%\textcolor{red}{Make sure we have cited some PRL papers.}\\
%\textcolor{red}{Make sure the manuscript follows all PRL guidelines of word limit, figures, etc..}

%\clearpage
\balance
%\bibliography{aipsamp}
\bibliography{main}% Produces the bibliography via BibTeX.

\clearpage
\section*{Supplement: Signal Model}

%\section{Signal Model}
Consider a RMTS-enhanced GPSM transmitter for a generic DP-MIMO system with $N_{T}$ Tx and $N_{R}$ Rx antennas. %Here, $N_{T} = N_{\textrm{SA}}$ denotes the number of Tx DP antennas, 
The RMTS is partitioned into $N_{\textrm{SA}}$ SAs connected to $N_{\textrm{RF}}=1$ RF chain. The transmitter employs a size-$P$ polarization constellation of $N_{\textrm{SA}} \times 1$ vectors $\mathbf{x}_0, \mathbf{x}_1,\dots,\mathbf{x}_p,\dots, \mathbf{x}_{P-1}$, each with an average symbol power ${\rm E}[|\mathbf {x}_p|^2]={1}$. We represent the polarization state of $\mathbf{x}_{p, n_{\textrm{SA}}}$ - the $n_{\textrm{SA}}$-th element of $\mathbf{x}_{p}$ - using Jones vector in a right-handed Cartesian coordinate system \cite{zhang2017polarization, zhang2019generalized} as
\begin{align}
    \mathbf{x}_{p, n_{\textrm{SA}}} = \frac{1}{\sqrt{N_{\textrm{SA}}}} 
    \begin{bmatrix}
    x_{p, n_{\textrm{SA}},{\rm V}} \\
    x_{p, n_{\textrm{SA}},{\rm H}} 
    \end{bmatrix} , \label{eqn:eqnSymbols}
\end{align}

where
\begin{align}
    x_{p, n_{\textrm{SA}},{\rm V}} = \cos \epsilon _{k, n_{\textrm{SA}}}\exp \left(\mathrm{j} 2 \pi \frac{m_{V, n_{\textrm{SA}}}-1}{M} \right), \\
    x_{p, n_{\textrm{SA}},{\rm H}} = \sin \epsilon _{k, n_{\textrm{SA}}}\exp \left(\mathrm{j} 2 \pi \frac{m_{H, n_{\textrm{SA}}}-1}{M} \right), \label{eqn:eqnPols}
\end{align}
are the signals transmitted by vertically and horizontally polarized transmit antennas, respectively; $\epsilon _{k, n_{\textrm{SA}}}$ denotes the polarization angle with $k=1,2,\ldots ,K$; and $m_{\mathrm {V}, n_{\textrm{SA}}}, m_{\mathrm {H}, n_{\textrm{SA}}}=1,2,\ldots ,M$ are the phase indices of the respective signals reflected by the $n_{\textrm{SA}}$-th SA. %Given the $M$-ary phase shift keying (PSK) and $K$ PolarSK signal constellations \cite{zhang2019generalized}, 
The set $\{M,K\}$ determines the size of the GPSM signal constellation so that $P=M^2K$. The signals with linear, circular or elliptic polarization states are determined by the parameter set $\{m_{V, n_{\textrm{SA}}}, m_{H, n_{\textrm{SA}}}, k\}$. %_{n_{\textrm{SA}}}\}$. %, which define the polarized signal reflected by the $n_{\textrm{SA}}$-th SA of the RMTS, where $m_{\mathrm{V}, n_{\mathrm{SA}}}$, $m_{\mathrm{H}, n_{\mathrm{SA}}}$, and $k_{n_{\mathrm{SA}}}$ denote the selection of $m_{\mathrm{V}}$, $m_{\mathrm{H}}$, and $k$ in the $n_{\mathrm{SA}}$-th activated transmit antenna, respectively. 
They are generated by each activated DP transmit SA according to (\ref{eqn:eqnSymbols}). For each transmission, $N_{\textrm{SA}}$ sets of $\{k$, $x_{p, n_{\textrm{SA}},{\rm V}}$, $x_{p, n_{\textrm{SA}},{\rm H}}\}$ are selected to convey $N_{\textrm{SA}} \log_{2}(M^2K)$ information bits. %because $N_{\textrm{SA}}$ polarized signals are transmitted by $N_{\textrm{SA}}$ DP SAs. 

Define the $N_{\textrm{SA}}\times 1$ vectors of the meta-atom reflection parameters $\boldsymbol {\theta}_{V,H}= [\theta _{1,V,H}, \cdots, \theta _{N_{\textrm{SA}},V,H}]^T$ and $\boldsymbol{\beta} = [\beta _{1,V,H}, \cdots, \beta _{N_{\textrm{SA}},V,H}]^T$, where $\theta_{n,V,H}\in [0, 2\pi)$ and $\beta_{n,V,H} \in [{0, 1}]$ denote the phase shift and the amplitude reflection coefficient for the $V$ and $H$ polarization components of the $n$-th SA of the RMTS, respectively; and $(\cdot)^T$ denotes the vector transpose. The $N_{\textrm{SA}}\times N_{\textrm{SA}}$ RMTS reflection-coefficient diagonal matrix is $\boldsymbol{\Theta}_{V,H} = \operatorname{diag}(\boldsymbol{\beta} \odot \exp [\mathrm{j} \boldsymbol{\theta}]) = \operatorname{diag} (\beta _{1,V,H}e^{\mathrm{j}\theta _{1,V,H}}, \cdots, \beta _{N_{\textrm{SA}},V,H} e^{\mathrm{j}\theta _{N_{\textrm{SA}},V,H}})$, where $\odot$ denotes the Hadamard product. By changing the bias voltage to each meta-atom of the RMTS, we tune $\theta_{n,V,H}$ and $\beta_{n,V,H}$ such that $\theta_{n_{\textrm{SA}},V,H}=\left(2 \pi \frac{-1+m_{V,H,n_{\textrm{SA}}}}{M} \right)$ to realize the polarization modulation states for $\mathbf{x}_{p, n_{\textrm{SA}}}$. 

The $2N_T \times 1$ transmitted symbol vector is 
$\mathbf{s} = [\mathbf{s}_{1}^T,...,\mathbf{s}_{n_T}^T,...,\mathbf{s}_{N_T}^T]^{T}$ where the modulated signal $\mathbf{s}_{n_T}$ is a two-element vector transmitted over the $n_{T}$-th dual-polarized SA antenna. The signals transmitted by $N_{\textrm{SA}}$ activated transmit SAs are denoted by the vector $\mathbf{x} = [\mathbf{x}_{p, 1}^T,...,\mathbf{x}_{p, n_{\textrm{SA}}}^T,...,\mathbf{x}_{p, N_{\textrm{SA}}}^T ]^T$. Note that $\mathbf{x}$ is a decimation of non-zero elements in $\mathbf{s}$ %\cite{zhang2019generalized} 
because $N_T>N_{\textrm{SA}}$. The conversion from low-dimension symbol vector $\mathbf{x}$ to high-dimension transmit signal $\mathbf{s}$ is achieved through a $2N_T \times N_{\textrm{SA}}$ index mapping matrix $\mathbf{U}$. Define $\mathcal{T} \subseteq \{1, 2, \cdots, 2N_T\}$ be the index set with $|\mathcal{T}| = N_{\textrm{SA}}$. Then, the $n$th row of matrix $\mathbf{U}$ is\par\noindent\small 
\begin{align}
[\mathbf{U}]_{n} = \begin{dcases}
\mathbf{0}^T_{N_{\textrm{SA}}},\;n\notin \mathcal{T},\\
\mathbf{1}^T_{N_{\textrm{SA}}},\;n\in\mathcal{T},
\end{dcases}
\end{align}\normalsize
where $\mathbf{0}_{N_{\textrm{SA}}}$ and $\mathbf{1}_{N_{\textrm{SA}}}$ denote $N_{\textrm{SA}}\times 1$ all-zeroes and all-ones vectors, respectively. Consequently, $\mathbf{s} = \mathbf{U}\mathbf{\Theta}_{V,H}\mathbf{x}$. 

Denote the polarization dependent channel path $\mathbf{H}_{n_{\mathrm{R}},{n_{\mathrm{T}}}} \in \mathbb{C}^{2\times 2}$ between each Tx subarray and receiver (Rx) pair as\par\noindent\small
\begin{equation} 
    \mathbf{H}_{n_{\mathrm {R}},{n_{\mathrm {T}}}} = \left [{ \begin{array}{ll} h_{\mathrm {VV},n_{\mathrm {R}},{n_{\mathrm {T}}}} & h_{\mathrm {VH},n_{\mathrm {R}},{n_{\mathrm {T}}}}\sqrt {\chi}\\ h_{\mathrm {HV},n_{\mathrm {R}},{n_{\mathrm {T}}}}\sqrt {\chi} & h_{\mathrm {HH},n_{\mathrm {R}},{n_{\mathrm {T}}}} \end{array} }\right ],
\end{equation}\normalsize
where $\chi$ denotes the expected power ratio of co-polarized channel to cross-polarized channel and each element denotes the channel gain from the ${n_{\mathrm {T}}}$-th DP antenna at the transmitter of a particular polarization (H or V) to the ${n_{\mathrm {R}}}$-th DP antenna at the Rx; the first polarization subscript corresponds to Tx. %Further EM analysis of $\chi$ is given in Eqn.~\ref{eqn:pcr}. 
We determine the value of $\chi$ through full-wave EM simulation (as shown in Fig.~\ref{fig:reflectionEMsim}). Then, the $2N_{R} \times 2N_{T}$ channel matrix $\mathbf{H}$ in terms of Tx subarray and Rx pairs $\mathbf{H}_{n_{\mathrm{R}},{n_{\mathrm{T}}}}$ is\par\noindent\small
\begin{align}
    \mathbf{H} = 
    \begin{bmatrix}
     \mathbf{H}_{1,1} & \mathbf{H}_{1,2} & \cdots & \mathbf{H}_{1,{N_{\mathrm{T}}}} \\
     \mathbf{H}_{2,1} & \mathbf{H}_{2,2} & \cdots & \mathbf{H}_{2,{N_{\mathrm{T}}}} \\
     \vdots & \vdots & \ddots & \vdots \\
    \mathbf{H}_{N_{\mathrm{R}},1} & \mathbf{H}_{N_{\mathrm{R}},2} & \cdots & \mathbf{H}_{N_{\mathrm{R}},{N_{\mathrm{T}}}}
    \end{bmatrix}.
\end{align}\normalsize

For line-of-sight (LoS) wireless communications, the standard statistical model for a multipath fading channel follows a Rician distribution. Here, the $N_{R} \times N_{T}$ complex channel impulse response (CIR) $\mathbf{H}$ is modeled as the sum of the fixed LoS component and a random multipath non-LoS (nLoS) channel component as \par\noindent\small
\begin{equation}
    \mathbf{H} = \sqrt{\frac{K_{c}}{K_{c}+1}} \mathbf{H}_{L} + \sqrt{\frac{1}{K_{c}+1}} \mathbf{H}_{N},
\end{equation}\normalsize
where $K_{c}$ is the real-valued Rician $K$-factor of the channel, $\mathbf{H}_{L} \in \mathbb{C}^{N_{R} \times N_{T}}$ is the LoS channel component that remains unchanged during the channel coherence time, and $\mathbf{H}_{N} \in \mathbb{C}^{N_{R} \times N_{T}}$ is the nLoS fading component representing random multipath fading.

The Rician $K_{c}$-factor is the power ratio of direct path (LoS) to other scattered nLoS paths. For a narrow-band block-fading channel, the received signal is %\par\noindent\small
\begin{equation}
    {\bf y}=\mathbf{H}\mathbf{s}+\mathbf{n}, \label{eqn:systemModel}
\end{equation}\normalsize
where ${\bf y}\in \mathbb{C}^{2N_{R}\times 1}$ is the output of $N_{R}$ Rx antennas, ${\bf s}\in \mathbb{C}^{2N_{T} \times 1}$ are the transmit symbols from $N_{T}$ dual-polarized transmit antennas, and ${\bf n}\in \mathbb{C}^{2N_{R}\times 1}$ is the circularly symmetric additive white Gaussian noise. 

The RMTS aperture dynamically reconfigures $N_{\mathrm {T}}$ by splitting it into multiple SAs. In full-aperture configuration, $N_{\mathrm {SA}} = N_{\mathrm {T}} = 1$. Given the RMTS aperture area $A$ and the operating wavelength $\lambda$, the full-aperture results

\clearpage
%\begin{comment}
%-------------------------------------------
%\begin{figure}[h!]
%  \centering
%  \includegraphics[width= 0.9 \linewidth]{fig02-new-v04-210228.PNG}
%  \caption{(a)An example of a RMTS-enhanced GPSM transmitter with a data rate of $R = 12$ bits/s/Hz mapping example bit sequences to their respective spatial and polarization states. The system parameters are $N_{T} = 4$, $N_{\textrm{SA}} = 2$, $K = 2$, and $M = 4$. The polarization of each transmitted signal calculated using Eqn.~\ref{eqn:eqnPols} is represented on the Poincar\'e sphere. In this bit frame, the symbol $\mathbf{s}_{A}=1 / \sqrt{2} \mathbf{x}_{k=1,q_{v}=3,q_{h}=4}$ is transmitted from SA $A$ and $\mathbf{s}_{C}= 1 / \sqrt{2} \mathbf{x}_{k=1,q_{v}=1,q_{h}=2}$ is transmitted from SA $C$. (b) The index bit mapping for this data frame and (c) the SA bit selector determine the activated SAs based on $l$.
%  }
%  \label{fig:modulationExample}
%\end{figure}
%-------------------------------------------
%\end{comment}
\onecolumngrid
%-------------------------------------------
\begin{figure*}%[ht!]
  \centering
  \includegraphics[width= 1.0 \textwidth]{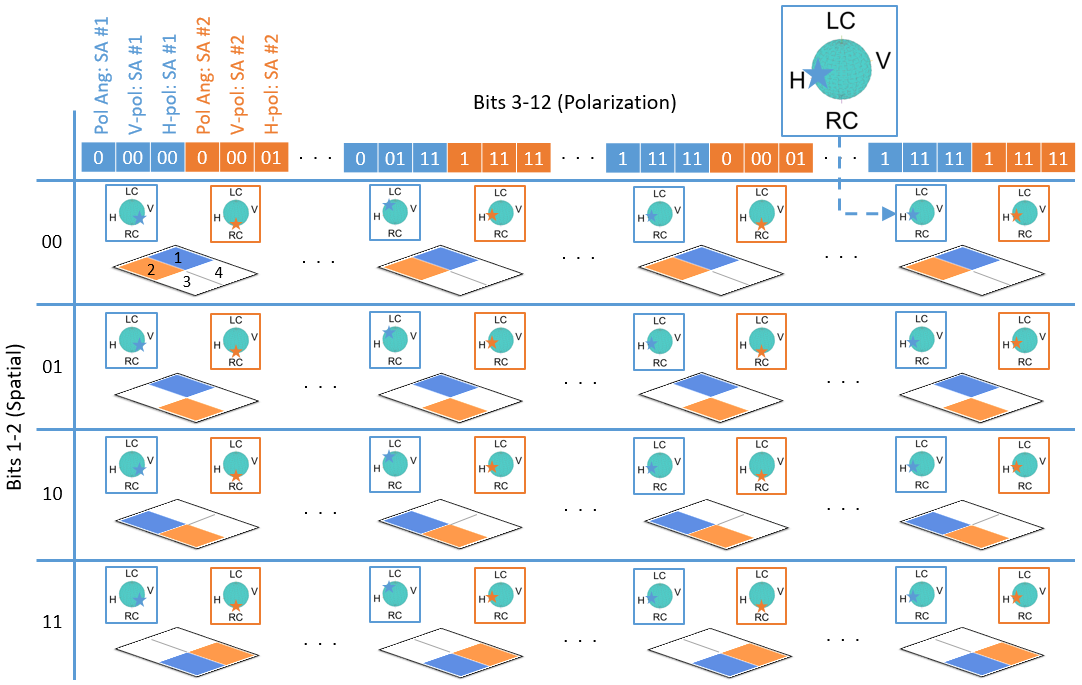}
  \caption{Illustration of mapping a 12-bit sequence to subarrays and polarization states in an RMTS-enhanced GPSM transmitter with data rate of $R = 12$ bits/s/Hz. Here, $N_{T} = 4$, $N_{\textrm{SA}} = 2$, $K = 2$, and $M = 4$. On the vertical axis, bits $1$-$2$ are spatial selector bits that determine the activated SAs. The colored elements are set to be active based on index bits. The indices of the SAs are labeled as $1$ to $N_{T}$ (upper left). On the horizontal axis, bits $3$-$7$ (blue) and $8$-$12$ (orange) determine the polarization states for the first and second activated SAs, respectively. The polarization of each transmit signal follows \eqref{eqn:eqnPols} and is represented on the Poincar\'e sphere (top right). 
  %\textcolor{red}{Caption and figure are good now. However, the spheres are not clearly visible. You may want to magnify one of the spheres in the white space on top.} %In this bit frame, the symbol $\mathbf{s}_{A}=1 / \sqrt{2} \mathbf{x}_{k=1,q_{v}=3,q_{h}=4}$ is transmitted from SA $A$ and $\mathbf{s}_{C}= 1 / \sqrt{2} \mathbf{x}_{k=1,q_{v}=1,q_{h}=2}$ is transmitted from SA $C$. (b) The index bit mapping for this data frame and (c) the SA bit selector determine the activated SAs based on $l$. \textcolor{red}{change the caption and figure}
  }
  \label{fig:modulationExample}
\end{figure*}
%-------------------------------------------
\twocolumngrid

\noindent in higher aperture directivity $D \approx 4 \pi A / \lambda^{2}$, increased Tx antenna gain and improved Rx SNR. 

When RMTS is configured as multiple sub-apertures (e.g. $N_{\mathrm {SA}} = 2$ or $4$), the data rate is increased by operating as a MIMO Tx at the expense of lower aperture directivity and gain (arising from reduced radiating area of each subarray as $A / N_{\mathrm {SA}}$).

For each transmission, the selections of polarization state and activated transmit DP antennas convey $N_{\mathrm {SA}}\lfloor \log _{2}(M^{2}K)\rfloor$ and $L=\left \lfloor{ \log _{2}\left ({\begin{smallmatrix} N_{\mathrm {T}} \\ N_{\mathrm {SA}} \end{smallmatrix}}\right )}\right \rfloor$ bits, respectively. The resulting SE is 
\begin{equation} 
    R=\left \lfloor{ \log _{2}\left ({\begin{smallmatrix} N_{\mathrm {T}} \\ N_{\mathrm {SA}} \end{smallmatrix}}\right )}\right \rfloor + N_{\mathrm {SA}}\lfloor \log _{2}(M^{2}K)\rfloor ~\mathrm {bits/s/Hz}.
\end{equation}

This is similar to conventional GPSM but no longer restricted by the physical $N_{\mathrm {RF}}$ (see Table~\ref{tbl:comparison}). At the Rx, the indices of combination of activated transmit antennas and transmit polarization state is determined for signal detection. After decoding the received symbols through, for example, the ML detector \cite{basar2016index}, BER is computed. 

Figure~\ref{fig:modulationExample} shows the index mapping procedure for this RMTS-enhanced GPSM transmitter. Data bits are divided into spatial- and polarization-signaling bits which determine the activated SAs and the transmit wave's polarization for each SA defined by  \eqref{eqn:eqnSymbols}-\eqref{eqn:eqnPols}. Denote the number of spatial-signaling bits by $q$. Then, the number of combinations of $N_{\textrm{SA}}$ transmitters that are selected by $q$ bits are $2^q = Q \le \left( \begin{array}{c} N_T \\ N_{\textrm{SA}} \end{array}\right)$. Denote the set of all $Q$ combinations of subarrays by $\mathcal{S}$, $|\mathcal{S}| = Q$. A bijective map $g:\{0,1\}^q \rightarrow \mathcal{S}$ maps the $q$ bits of the codeword to activated subarrays in the set $\mathcal{S}$. For example, when $q=2$, $N_{SA} = 2$, $N_{T} = 4$ with each transmitter indexed by an element of the set $\{1,2,3,4\}$, and $\mathcal{S} = \{(1, 2), (1, 3), (2, 3), (3, 4)\}$, a possible mapping of spatial bits to activated SAs is
\begin{align}
        \underbrace{\{00, 01, 10, 11\}}_{\text{Spatial bits}} \xrightarrow[]{g(\cdot)} 
        \underbrace{\{(1, 2), (1, 3), (2, 3), (3, 4)\}}_{\text{Active SAs}}.\nonumber
\end{align}

\end{document}